\documentclass[%
preprint,
superscriptaddress,
 amsmath,amssymb,
 aip,
 reprint,
onecolumn
]{revtex4-1}

\usepackage{graphicx}
\usepackage{dcolumn}
\usepackage{bm}
\usepackage [latin1]{inputenc}
\usepackage{amsmath}
\usepackage{epstopdf}
\usepackage{graphicx}
\usepackage{hyperref}
\usepackage{color}
\usepackage[english]{babel}
\usepackage{lmodern}
\usepackage[T1]{fontenc}
\usepackage[normalem]{ulem}
\usepackage{color}
\usepackage{lineno}

\makeatother

\newcommand\OM{\Omega}

\newcommand{\be}{\begin{equation}}
\newcommand{\ee}{\end{equation}}
\newcommand{\bea}{\begin{eqnarray}}
\newcommand{\eea}{\end{eqnarray}}

\begin{document}


\title{Lattice Boltzmann Multicomponent Model for Direct-Writing Printing}

\author{Michele Monteferrante}
\thanks{First author}%
\affiliation{Istituto per le Applicazioni del Calcolo CNR, Via dei Taurini 19, 00185 Rome, Italy}
\author{Andrea Montessori}
\affiliation{Istituto per le Applicazioni del Calcolo CNR, Via dei Taurini 19, 00185 Rome, Italy}
\author{Sauro Succi}
\affiliation{Istituto per le Applicazioni del Calcolo CNR, Via dei Taurini 19, 00185 Rome, Italy}
\affiliation{Center for Life Nano Science$@$La Sapienza, Istituto Italiano di Tecnologia, 00161 Roma, Italy}
\affiliation{Harvard Institute for Applied Computational Science, Cambridge, Massachusetts, United States}
\author{Dario Pisignano}
\affiliation{Dipartimento di Fisica, Universit\`a di Pisa, Largo B. Pontecorvo 
16
 3, 56127 Pisa, Italy}
\affiliation{NEST, Istituto Nanoscienze-CNR, Piazza S. Silvestro 12, 56127 Pisa, Italy}
\author{Marco Lauricella}
\email[Corresponding author: ]{m.lauricella@iac.cnr.it}
\affiliation{Istituto per le Applicazioni del Calcolo CNR, Via dei Taurini 19, 00185 Rome, Italy}


\begin{abstract}

We introduce a mesoscale approach for the simulation of multicomponent flows to model the direct-writing printing process, along with the early stage of ink deposition.
As an application scenario, alginate solutions at different concentrations are numerically investigated alongside processing parameters, such as apparent viscosity, extrusion rate, and print head velocity.
The present approach offers useful insights on the ink rheological effects upon printed products, susceptible to geometric accuracy and shear stress, by manufacturing processes such as the direct-writing printing for complex photonic circuitry, bio-scaffold fabrication, and tissue engineering.
\end{abstract}

\maketitle


\section{Introduction}\label{sec:01}

In the last decade, 3D printing processes have gained enormous attention as tool for additive manufacturing in many fields of science and engineering. 
The major success of 3D printing is mostly due to the digital process control, which offers remarkable flexibility in terms of patterning through material deposition. 
By exploiting a computer-controlled layer-by-layer fabrication technique, soft materials are utilised in fused deposition modelling and 
in extrusion direct-writing bio-plotters that apply a pressure gradient to fluids, possibly generating architected matter with qualitatively new properties \cite{truby2016printing,ngo2018additive}. 
As a consequence, this set of technologies is nowadays exploited in a wide variety of applications, such as tissue engineering (e.g., bio-compatible scaffolds), microsystems (lab-on-chip), microelectronics (sensors), and aerospace structures (aircraft engine bracket), to name a few \cite{farahani2016three,axpe2016applications,utela2008review}. 
The vast potential of additive manufacturing requires, however, an unprecedented control over several aspects of the soft materials involved in the 3D printing process. 
Their dynamics, composition, structure, function and rheology are all key elements, which severely affect the features of the finally produced parts and structures.

In this framework, efficient numerical simulations might offer a crucial help to understand the relevant, and interplaying, characteristics 
of the fluids and experimental setups, similarly to other manufacturing processes where computational tools have been successfully applied in the last two decades (e.g. electrospinning \cite{lauricella2020models}, electrospray \cite{ganan2018review}).
In the direct-writing printing context, numerical simulations could be used to maximise the printability of a given ink, avoid process failures and anticipate the 
microstructural properties of the products, by providing a set of observables (e.g., flow rate, stress tensor), which are often 
difficult to access experimentally. 
Printability is usually studied empirically, and \textit{a priori} criteria are not available that, given an ink and a prototype model, allow 
the success of the manufacturing process to be reliably predicted. 
Moreover, the challenging characteristic length scale of the process, \textit{i.e.} the diameter of the print head nozzle 
used to deposit materials, significantly constrain the possible choices within available computational methods. 
For instance, microscopic techniques, such as molecular dynamics, are generally unable to access length 
and time scales of experimental relevance, for want of computing resources. 
On the other hand, finite volume or finite element methods may also become computationally expensive
in the presence of moving boundary conditions, such as the ones describing the moving print head.

Mesoscale techniques, and particularly the Lattice Boltzmann (LB) method \cite{succi2018lattice,kruger2017lattice,huang2015multiphase} 
offer an appealing alternative to both methods above, eventually striking an optimal balance between the two.
Indeed LB moves noise-free discrete probability distributions along force-free (straight) trajectories 
and represents the effect of molecular collisions through the relaxation towards a suitable lattice local-equilibrium.
Once the lattice symmetry and the local equilibria are suitably designed, the scheme can be shown to 
reproduce quantitatively the Navier-Stokes equations of fluid flows.
The result is a very elegant and efficient computational scheme, featuring outstanding 
amenability to parallel implementations also in the presence of strong nonlinearities and complex boundary conditions
\cite{benzi1992lattice,succi2018lattice,kruger2017lattice}.

In this work, we open a new route for predicting 3D printability, developing the regularised version of the 
Colour Gradient (CG) Lattice Boltzmann (LB) model \cite{succi2018lattice,kruger2017lattice,huang2015multiphase} to account for the non-Newtonian rheological behaviour, typical of 3D printed pseudo-plastic inks. 
These systems endure a largely varying apparent viscosity, depending on the shear rate \cite{cross1965rheology}. 
The regularised version of the LB method mitigates issues related to both low and high viscosities \cite{latt2006lattice}, the first 
threatening numerical stability, while the latter undermining the very hydrodynamic limit of the LB scheme. 
Further, the shift in the nozzle position during actual 3D printing processes is also included. 
As a practical application, we focus on sodium  alginate solutions, which are widely used in direct-writing printing 
to manufacture scaffolds for cell cultures and tissue regeneration \cite{axpe2016applications}.
The printing accuracy is discussed in terms of a Parameter Optimization Index (POI) \cite{webb2017parameter}, which is predicted
in terms of the numerical inputs.

\section{Model details}\label{sec:01}
\subsection{Regularized colour-gradient lattice Boltzmann model}\label{sec:01A}

The regularized CG method for multicomponent-multiphase systems provides computationally efficient access to capillary number regimes 
relevant to microfluidics \cite{montessori2018regularized}. 

The general form of CG LB equations \cite{leclaire2017generalized,huang2015multiphase,leclaire2013progress} writes as follows:  
\be
f^k_i(\vec x+\vec c_i \Delta t,t+\Delta t)=f^k_i(\vec x,t)+\Omega^k_i[f^k_i(\vec x,t)], 
\label{eq:lbm}
\ee
where $k$ is the colour component, $i$ is the index spans the lattice discrete directions, and $\Omega^k_i$ denotes the collision operator of the Colour Gradient model. 
In Eq. \ref{eq:lbm}, $f_{i}^{k}$, represents the probability of finding a particle of the $k-th$ component at position 
$\vec{x}$ and time $t$ with discrete lattice velocity $\vec{c}_{i}$. For the sake of simplicity, we adopt the standard D2Q9 lattice, and only two colours are assumed in the system, \textit{i.e.} yellow and blue, standing for the dense ink and air, respectively. 
In the following, $i=0$ denotes the resting population with zero velocity, while $i \in [1, \dots, 8]$ represent the directions at angle $\theta=(i-1)\pi/4$ with velocity modulus $|\vec{c}_i|=1$ for $i \in [1,3,5,7]$, and $|\vec{c}_i|=\sqrt{2}$ for $i \in [2,4,6,8]$ in lattice units, assumed $\Delta x_{\text{LB}}=1$ and  $\Delta t_{\text{LB}}=1$. 
A comparison among the regularized version of the Colour Gradient (CG) lattice Boltzmann (LB) model \cite{montessori2018regularized} and other LB diffuse interface approaches (e.g., pseudo-potential model, free energy model) was discussed by S. Leclaire and coworkers \cite{leclaire2017three} noting that, although both pseudo-potential and CG methods are able to reproduce the correct physics, macroscopic parameters such as surface tension and interface thickness can be set independently only in CG method, while pseudo-potential model needs {\it a posteriori} adjustment of the repulsive forces between different species to match the required physics.
The density, $\rho_{k}$, of the $k-th$ fluid component is assessed as the zeroth moment of the distribution functions
\begin{equation}
\rho^{k}\left(\vec{x},\,t\right) = \sum_i f_{i}^{k}\left(\vec{x},\,t\right),
\end{equation}
while the total momentum, $\rho \vec{u}$, is defined by the first order moment:
\begin{equation}
\rho \vec{u} = \sum_k \sum_i   f_{i}^{k}\left(\vec{x},\,t\right) \vec{c}_{i},
\end{equation}
being $\rho$ the sum of the two component densities. 
The collision operator, $\Omega^k_i$, results from the combination of three sub-operators: 
\be
\label{eq:cgoperator}
\OM^k_i=(\OM^k_i)^{(3)}[(\OM^k_i)^{(1)}+(\OM^k_i)^{(2)}].
\ee
The first, $(\OM^k_i)^{(1)}$, is the standard BGK operator:
\be\label{eq:modbgk}
(\OM^k_i)^{(1)}=-\frac{1}{\tau} [f^k_i(\vec x,t)-f^{k,eq}_i(\vec x,t)],
\ee
where $\tau$ is a relaxation time setting the numerical viscosity of the mixture (see below) and $f^{k,eq}_i(\vec x)$ is a modified equilibrium distribution function: 
\bea
&& f^{k,eq}_i(\vec x,\rho_k,\vec{u})=\nonumber\\
&& \rho^k\Big( \phi^k_i+w_i \Big[ \frac{\vec c_i\cdot\vec{u}}{c_s^2}+\frac{(\vec c_i\cdot \vec{u})^2}{2 c_s^2}-\frac{(\vec{u})^2}{2 c_s^2}\Big]\Big),\nonumber\\
\eea
with $c_s$ the lattice sound speed and $w_i$ the weights of the standard D2Q9 lattice \cite{huang2015multiphase}: $w_0=4/9$, $w_{1,3,5,7}=1/9$, $w_{2,4,6,8}=1/36$.
Here, the coefficients, $\phi^k_i$, read \cite{leclaire2013progress}:
\be
\phi_{i}^{k}=\begin{cases}
\alpha_{k}, & i=0,\\
\left(1-\alpha_{k}\right)/5, & i=1,3,5,7,\\
\left(1-\alpha_{k}\right)/20, & i=2,4,6,8,
\end{cases}
\ee
and are tuned to simulate systems with different density ratio $\gamma$:
\bea
\gamma={\rho_Y\over\rho_B}={1-\alpha_B\over 1-\alpha_Y},
\eea
with the apexes $Y$ and $B$ standing for yellow and blue fluid component, respectively. The partial pressure of $k-$th component reads:
\bea
\label{eq:cgoef}
p^k={3\over 5}\rho^k(1-\phi^k_0).
\eea
The second operator, $(\OM^k_i)^{(2)}$, called perturbation operator, generates the interfacial tension and has the form:  
\be\label{eq:pertopform}
(\OM^k_i)^{(2)}={A\over2}|\nabla \rho^N|\Big[w_i{(\vec{F}_{\text{cg}} \cdot \vec c_i)^2\over |\vec{F}_{\text{cg}} |^2}-B_i\Big],
\ee
where $\vec{F}_{\text{cg}}$ denotes the colour gradient force, reading:
\be
\vec{F}_{\text{cg}}=\frac{\rho_B}{\rho} \nabla \Big(\frac{\rho_Y}{\rho} \Big) -\frac{\rho_Y}{\rho} \nabla \Big(\frac{\rho_B}{\rho} \Big).
\ee
As observed in Refs \cite{wen2019improved,saito2017lattice,liu2012three}, the gradient for an arbitrary observable $\chi$ can be obtained by the second-order isotropic central scheme :
\bea\label{eq:lattgrad}
\nabla \chi(\vec x,t)={1\over c_s^2}\sum_{i} w_i \chi(\vec x+\vec c_i,t)\;\vec c_{i}
\eea
In Eq. \ref{eq:pertopform}, the $B_i$ coefficients depend on the lattice (taken: $B_0=-4/27$, $B_{1,3,5,7}=2/27$, $B_{2,4,6,8}=5/108$ from Ref. \cite{leclaire2013progress}), whereas $A$ is a free parameter modeling the surface tension, $\sigma$, that is:\cite{leclaire2013progress,reis2007lattice}: 
\be
\sigma={2 \tau \over 9} A ,
\ee
where $\tau$ is the effective relaxation time.
The recoloring operator $(\OM^k_i)^{(3)}$ is necessary since the perturbation operator alone does not guarantee the phase separation:
\bea
(\OM_i^Y)^{(3)}={\rho_Y\over\rho}f_i^*+\beta{\rho_Y\rho_B\over\rho^2}\cos(\varphi_i) \sum_k f_i^{eq}(\vec x,\rho_k,\vec{u}=0)\\
(\OM_i^B)^{(3)}={\rho_B\over\rho}f_i^*-\beta{\rho_Y\rho_B\over\rho^2}\cos(\varphi_i) \sum_k f_i^{eq}(\vec x,\rho_k,\vec{u})
\eea
Here, $\beta$ is a parameter tuning the thickness of the diffuse interface, $f_i^*$ is the post collision total density in the lattice direction $i$ , $f_i^{eq}=\sum_k f_i^{k,eq}$ and finally:
\be
\cos(\varphi_i)={{\vec c}_i\cdot \nabla \rho_N\over |{\vec c}_i||\nabla \rho_N| }.
\ee
The kinematic viscosity, $\nu$, is assessed as \cite{leclaire2017generalized,leclaire2013progress}:
\be\label{eq:tauinter}
\frac{1}{\nu} =\frac{\rho_Y}{\rho} \frac{1}{\nu_Y}+\frac{\rho_B}{\rho}\frac{1}{\nu_B},
\ee
being $\nu_Y$ and $\nu_B$ the density of the yellow and blue component, respectively.
In order to model the wettability on the different walls in the system, see Fig.~\ref{fig:boundaries}, and compute the gradients of $\rho^k$ by Eq.~\ref{eq:lattgrad} also close to the boundaries, we set a fictitious fluid density, $\rho^k_s$, for the two components on all the wall nodes \cite{latva2005static}. The fictitious densities are estimated by the extrapolation of the color function at neighboring fluid lattice nodes by the formula:
\be\label{eq:wet}
\rho^k_s(\vec x,t) = \zeta^k(\vec x,t)  \frac{\sum_{i} w_i \rho^k(\vec x+\vec c_i,t)}{\sum_{i} w_i} s(\vec x+\vec c_i,t),
\ee
where $\zeta^k(\vec x,t)$ is a positive parameter tuning the affinity of the different walls, see Fig.~\ref{fig:boundaries},  at the position $\vec x$ for a given fluid component, and $s$ is a switch function taking value one if the site at $\vec x+\vec c_i$ is a fluid and is zero otherwise. The contact angle is given by:
\bea\label{eq:wet2}
\theta=\arccos\Big({\rho_{sB}-\rho_{sY}\over\rho_B^0}\Big)
\eea
where $\rho_B^0$ is the initial density of the blue fluid.
Note the present strategy can be interpreted as a simplified version of the approaches reported in Refs \cite{akai2018wetting,leclaire2016modeling} where the wall densities are interpolated with the same expression of Eq.~\ref{eq:wet} without the $\zeta^k$ but the density's gradient is subsequently rotated to match the prescribed contact angle. Although less accurate in reproducing the contact angle for the presence of spurious currents, our approach is a simple procedure to model hydrophobicity ($\zeta^k<1$) or hydrophilicity ($\zeta^k>1$) of the walls as given in Fig.~\ref{fig:boundaries}. 

Implying the Einstein convention for summation over Greek indices (see Appendix A of Ref. \cite{kruger2017lattice}), the regularization step reads \cite{montessori2018regularized}:
\be
f^{k, reg}_i(\vec x,t)=f^{k,eq}_i(\vec x,\rho_k,{\bf u})+\frac{w_i}{2 c_s^4}Q_{i \alpha \beta} \Pi^{neq, k}_{\alpha \beta}, \label{eq:reg}
\ee
where $Q_{i \alpha \beta}=(c_{i \alpha} c_{i \beta}-c_s^2 \delta_{\alpha \beta})$ and $ \Pi^{neq, k}_{\alpha \beta}=(\sum_i f_i^k c_{i \alpha} c_{i \beta}) - (\sum_i f_i^{k, eq} c_{i \alpha} c_{i \beta})$ with $\alpha, \beta$ denoting Cartesian directions and $\delta$ the Kronecker delta.
Note that Eq. \ref{eq:reg} consists of a projection of a distribution functions, $f_i^k$, onto the set of Hermite basis. In doing so, we obtain a set of filtered distribution functions, $f^{k, reg}_i$, which depends only on the first and second macroscopic hydrodynamic moments without higher-order non-equilibrium information often referred as ghost moments \cite{montessori2018regularized,latt2006lattice,zhang2006efficient}. It was shown \cite{coreixas2018high,montessori2014regularized,latt2007hydrodynamic} that the procedure provides general benefits in terms stability in the BGK LB scheme, which can be decisive in the case of low-viscosity simulations. Hence, the regularized distributions, $f^{k, reg}_i$ are used in Eq. \ref{eq:lbm}.

The hydrodynamic limit of Eq. \ref{eq:lbm} \cite{huang2015multiphase,reis2007lattice} is found to converge to a set of equations 
for the conservation of mass and linear momentum:
\bea
\frac{\partial \rho}{\partial t} + \nabla \cdot {\rho \vec{u}}=0 \:\\
\frac{\partial \rho \vec{u}}{\partial t} + \nabla \cdot {\rho \vec{u}\vec{u}}=-\nabla p + \nabla \cdot [\rho \nu (\nabla \vec{u} + \nabla \vec{u}^T)] +\nonumber\\ +\nabla \cdot \bm{\Sigma} \:
\eea
where $p=\sum_k p_k$ is the pressure, $\nu=c_s^2(\tau-1/2)$ is the kinematic viscosity of the mixture, $\bm{\Sigma}=-\tau\sum_k \sum_i\left(\Omega_{i}^{k}\right)^{(2)} \vec{c}_i \vec{c_i}$ is the stress tensor of the curved interface, and $\tau$ is a time tuning the relaxation of the fluid flow towards its local equilibrium used in the collision operator, $(\OM^k_i)^{(1)}$, of Eq. \ref{eq:modbgk}. At each time step before the collision in Eq. \ref{eq:lbm}, all the populations, $f_i^k$, are filtered by applying the regularization step \cite{montessori2018regularized,latt2006lattice}.

It is worth to highlight that the LB approach avoids two potential and serious issues of computational physics in discretizing the Navier-Stokes equations of continuum fluid mechanics: strong non-linearity and complex geometry within a time-dependent formulation. 
In particular, the discretization of the non-linear term, $\nabla \cdot {\rho \vec{u}\vec{u}}$, in the Navier-Stokes equations requires the non-locally derivative approximations over the adjacent space in numerical techniques such as finite-difference methods and finite element methods. In contrast, the LB method disentangles the non-locality and non-linearity of the problem, since the non-linearity is treated locally (collision step of Eq. \ref{eq:lbm}), and the non-locality is treated linearly (streaming step of Eq. \ref{eq:lbm}) as a shift in memory over the adjacent nodes. Thus, it turns out that the LB approach is a very attractive computational bargain to high-performance computing on parallel architectures, including GPUs \cite{succi2018lattice,kruger2017lattice}.

\subsection{Extension to non-Newtonian flow and moving print head}\label{sec:01B}

To model the non-Newtonian fluids, the model is extended in similarity with the
approach reported in Refs \cite{lauricella2018entropic,pontrelli2009unstructured,gabbanelli2005lattice,aharonov1993non}. 
The extension consists essentially of determining the local value of the relaxation time, in such a way that the desired local value of the viscosity is recovered to match a constitutive equation for the stress tensor \cite{malaspinas2007simulation,ouared2005lattice,gabbanelli2005lattice,aharonov1993non}.
We assume that the shear-thinning model introduced originally by M. Cross \cite{cross1965rheology} (in the following referred to as Cross model) adequately describes the ink viscosity. Note that the Cross model was already employed to describe the non-Newtonian behavior of alginate solutions by Roopa and Bhattacharya \cite{roopa2009characterisation}. However, other possible models can be freely adopted without losing the applicability of the present approach. 
The Cross model states:
\be\label{eq:notnewvisc}
\mu(\dot \gamma)=\mu_{\infty}+{\mu_0-\mu_{\infty}\over 1+(\lambda\dot\gamma)^n},
\ee
where $\mu$ is the dynamic viscosity, $n$ the flow index ($n<1$ for a pseudoplastic fluid), $\mu_0$ the zero shear viscosity, $\mu_{\infty}$ the asymptotic value, and $\lambda$ the retardation time at which the shear-thinning starts.
In the following, the yellow dense component (the ink) is assumed to show a non-Newtonian behaviour \cite{cross1965rheology} while the blue component (the air) is a Newtonian fluid. Far from the interface, the stress tensor and the strain tensor are mainly represented by the k-th component so that $\Pi_{\alpha \beta}\sim \Pi_{\alpha \beta}^k$ and $\Gamma_{\alpha \beta}\sim\Gamma_{\alpha \beta}^k$, respectively.
Following the literature \cite{succi2018lattice,ouared2005lattice}, the stress tensor $\Pi_{\alpha \beta}$ relates with the strain tensor $\Gamma_{\alpha \beta}$ by the relation
$\Gamma_{\alpha \beta}=-(1/2\rho \tau c^2_s)\Pi_{\alpha \beta}$, where the stress tensor $\Pi_{\alpha \beta}= \sum_{i}\left(f_i - f_i^{ eq} \right)\vec{c}_{i \alpha} \vec{c}_{i \beta}$. 
Thus, in the yellow fluid bulk the last relation can be rewritten as:  
\be\label{eq:stressvsstrain}
\dot \gamma_Y(|\Pi^Y_{\alpha\beta}|)={|\Pi^Y_{\alpha\beta}|\over \rho_Y\tau_Y(\dot\gamma_Y)c_s^2},
\ee
where the stress tensor of the yellow component reads $\Pi_{\alpha \beta}^Y= \sum_{i}\left(f_i^Y - f_i^{Y,eq} \right)\vec{c}_{i \alpha} \vec{c}_{i \beta}$, and the tensor norms are computed as $\dot\gamma_Y=2\,|\Gamma_{\alpha\beta}^Y|=2\sqrt{\sum_{\alpha\beta}\Gamma_{\alpha\beta}^Y\Gamma_{\alpha\beta}^Y}$ and $|\Pi_{\alpha\beta}^Y|=\sqrt{\sum \Pi_{\alpha\beta}^Y\Pi_{\alpha\beta}^Y}$ with the relaxation parameter $\tau_Y$ setting the kinematic viscosity of the yellow fluid, $\nu_Y=c_s^2(\tau_Y -1/2)$.

Since $\mu(\dot\gamma)=\rho\nu(\dot\gamma)$ and $\mu(\dot\gamma)=\rho c_s^2(\tau(\dot\gamma)-1/2)$, the function $\tau_Y(\dot\gamma_y)$, requested in Eq. \ref{eq:stressvsstrain}, can be obtained by Eq. \ref{eq:notnewvisc} as:
\be\label{eq:tauofgammadot}
\tau_Y(\dot\gamma_Y)={1\over 2}+{1\over c_s^2}[\nu_{\infty, Y}+{\nu_{0, Y}-\nu_{\infty, Y}\over 1+(\lambda\dot\gamma_Y)^n}].
\ee

Inserting Eq. \ref{eq:tauofgammadot} in Eq. \ref{eq:stressvsstrain} provides an implicit equation in the variable $\dot\gamma_Y$, which is solved iteratively, performing several iterations as long as the value of $\dot\gamma_Y$ is not converged below a given threshold. If close to the interface, $\tau$ is computed from the interpolated value of the viscosity by Eq.~\ref{eq:tauinter}. A similar approach was exploited by Pontrelli et al. \cite{pontrelli2009unstructured} to model a pseudo-plastic single-phase fluid, and it was validated by comparison with the analytical solution of Poiseuille flow with the power-law model.



Since the print head moves during the  process, we needed a particular treatment of the boundary conditions of the nozzle walls and fluid nodes around the nozzle. Inspired to the trailblazing work by Antony Ladd \cite{ladd1994numerical1}, we define a template of solid nodes with an internal reference system, which is translated along the time evolution by integrating an equation of motion. In accordance with the formulation proposed by F. Jansen and J. Harting \cite{jansen2011bijels}, only the exterior regions are filled with fluid, whereas the interior parts of the nozzle is considered solid nodes. 

Denoting with $f_i^*(\vec x_b,t)$ the post collisional distribution at the boundary position $\vec x_b$ hitting the nozzle wall  in the direction $\vec c_i$ and located in the middle position along the link connecting the solid node $\vec x_s$ from the boundary fluid node  $\vec x_b$, we exploit a simple generalization of the halfway bounce-back rule \cite{kruger2017lattice,ladd2001lattice,ladd1994numerical1}. Hence, the streaming step proceeds as:
\be
f^k_{\bar i}(\vec x_b,t+1)=f^k_i(\vec x_b,t)-2w_i\rho^k_w \frac{\vec{\upsilon}_{\text{nozzle}}\cdot\vec c_i}{c_s^2}
\label{eq:movwall}
\ee
where $\bar i$ is the lattice direction $-\vec c_i$. The symbol $\rho^k_w$ in Eq.~\ref{eq:movwall} denotes the density at the wall position, which is obtained by a third order interpolation in the direction $-\vec c_i$.  

Because the print head moves over the lattice nodes, it happens that a subset of fluid boundary nodes in front of the moving nozzle crosses its surface, becoming solid nodes. Similarly, a subset of interior nodes on the surface is released at the back of the nozzle. The two distinct events require the destruction and the creation of fluid nodes, respectively. Following the previous strategy reported in the literature, whenever a fluid node changes to solid, the fluid is deleted \cite{aidun1998direct} without transferring its linear momentum to the nozzle beneath the nozzle infinite mass hypothesis.
In the creation fluid node step, as the nozzle leaves a lattice 
site, new fluid populations are initialized from the equilibrium distributions,
$f^{eq,k}_i(\bar{\rho}^k,\vec{\upsilon}_{\text{nozzle}})$, for the two $k$-th components with the velocity of the nozzle wall, $\vec{\upsilon}_{\text{nozzle}}$ and the $k-$th fluid density taken as its average value, $\bar{\rho}^k$, computed over the neighbouring  fluid nodes \cite{aidun1998direct,jansen2011bijels}.

As a first approximation, the solid-fluid interactions are accounted only 
for the part concerning the effect of the moving nozzle on the surrounding 
fluid and not viceversa, which is equivalent to assume that the motion of 
the print head is fully controlled by the digital system of the 3D printer 
(nozzle infinite mass hypothesis). 
Hence, a constant velocity $\vec{\upsilon}_{\text{nozzle}}$ of the nozzle 
(print head) is taken as an input parameter to describe the linear motion 
of the nozzle, and a drag force is added close to the deposition zone in 
order to model the friction between the ink and the collector. 

Inside the nozzle, the yellow component (the ink) is inserted with constant velocity $\vec{u}_{\text{ink}}$ with respect to the internal reference frame on the print head. Considering that the nozzle reference frame is moving with $\vec{\upsilon}_{\text{nozzle}}$, the total fluid velocity inserted at the inlet reads: 
\be
\vec{u}_{\text{Inlet}}=\vec{u}_{\text{ink}}+\vec{\upsilon}_{\text{nozzle}}.
\label{eq:movnozzle}
\ee
Hence, $\vec{u}_{\text{Inlet}}$ replaces $\vec{\upsilon}_{\text{nozzle}}$ in Eq. \ref{eq:movwall}, which is used to model the fluid inlet inside the print head.

We also used a Dirichlet boundary conditions in our work to maintain the pressure (density) of the blue fluid (air) constant. For the Dirichlet condition, the anti-bounce back scheme \cite{ginzburg2005generic} is used instead for constant pressure (densities) boundaries:
\bea
f^k_{\bar i}(\vec x_b,t+1)=-f^{k*}_i(\vec x_b,t)+\nonumber\\
+2 \rho_w^k\Big\{ \phi^k_i+w_i\Big[ \frac{(\vec c_i\cdot\vec{u}_w)^2} {2c_s^4} -\frac{\vec{ u}^2_w}{2c_s^2}\Big]\Big\},
\eea
where $\rho_w^k$ is the imposed density at the open boundary and $\vec{u}_w$ is the velocity at half-way point estimated by a second order interpolation along the direction $-\vec{c}_i$.

The drag force modelling the friction between the ink and the collector reads:
\be
F_{d}(\vec x,t)=-\gamma\rho_Y(\vec x,t)\vec{u}(\vec x,t),
\ee
where $\gamma$ is the friction coefficient tuning the drag force. This force is turned on at grid points which are closer than 4 lattice units from the deposition wall. The drag force is added in Eq. \ref{eq:modbgk} by the exact difference method proposed by Kupershtokh and coworkers \cite{kupershtokh2009equations}.

\section{System Setup}\label{sec:02}

It is worth to remark the main simplifying assumptions adopted in the simulations.
In the present paper, the Cross model is adopted to describe the non-Newtonian fluid, albeit any other rheological relation can be adopted, obviously in the context of pseudo-plastic models (e.g., Carreau Model, Herschel-Bulkley model, etc.). Further, the adhesion property of the fluid onto the deposition surface can be relevant. The contact angle is set equal to 90$^\circ$, corresponding to the neutral affinity of the ink to the surface (neither hydrophilic nor hydrophobic). Nonetheless, different contact angle values can be investigated by the present model. Finally, we exploit a two-dimensional description of the system to reduce both the computational cost and the wall-clock time necessary for each simulation. As a consequence, the lateral shear effect of the slender fluid filament along the third dimension is neglected in the two-dimensional description. Nonetheless, the comparison with experimental data in Section \ref{sec:04} will show as the two-dimensional approximation does not invalidate the agreement between the numerical results and the experimental counterpart. In other words, the validation of the present model is not prevented from the dimension reduction.

\begin{figure}[!hbtp]
\includegraphics[width=0.7\hsize,clip]{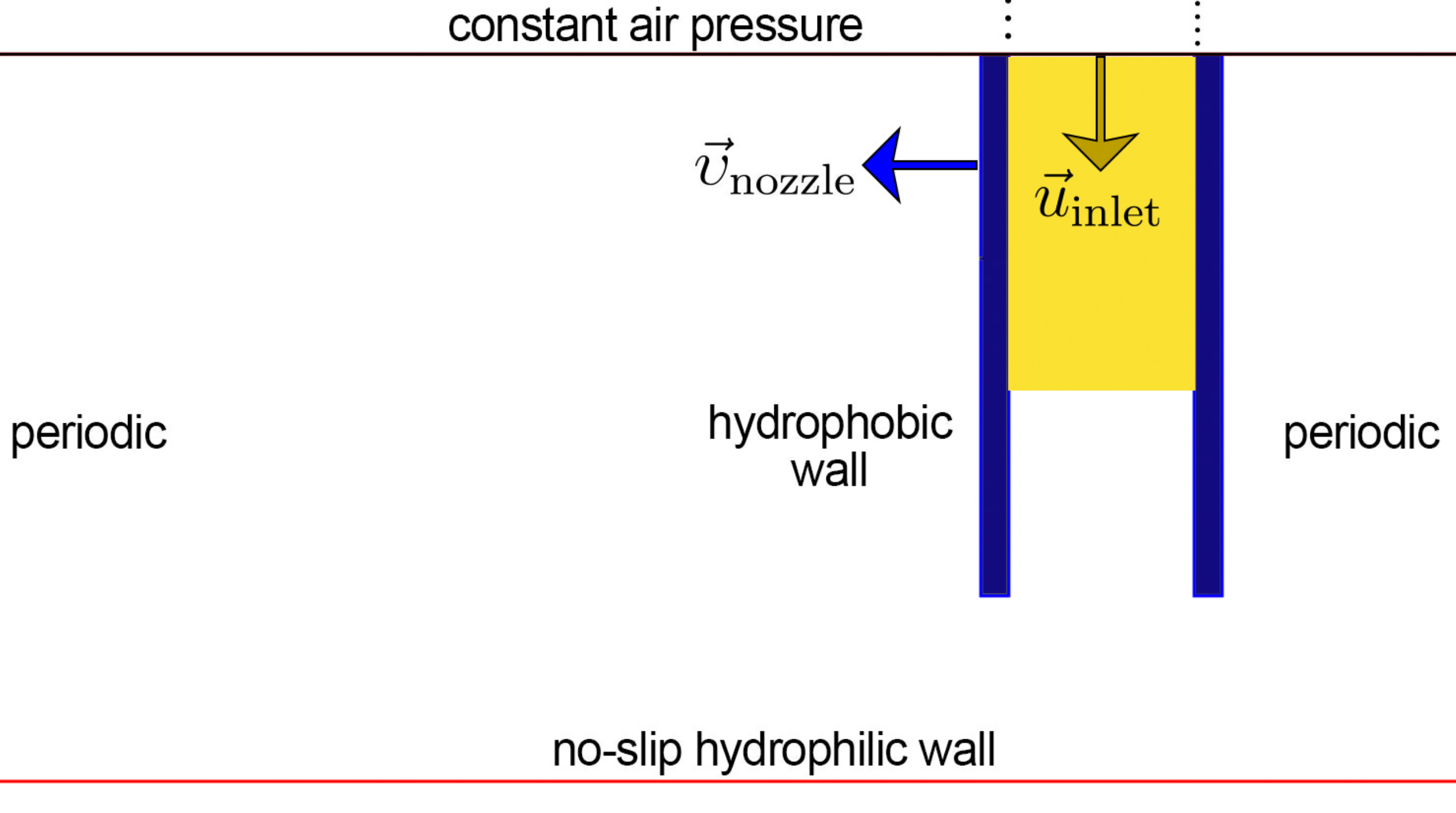}
\caption{Representation of the system alongside with the different treatment of boundary conditions.}
\label{fig:boundaries}
\end{figure}

The alginate concentration in water is taken in the range $1\%-3\%$ w/v, 
allowing the investigation of the fluid characteristics for optimal printing.
Alginate solutions are shear-thinning non-Newtonian fluids 
\cite{sarker2017modeling,roopa2009characterisation}. 
Thus, the viscosity decreases to smaller value as the shear rate increases. 
This dependence of the viscosity on the shear rate makes the whole printing 
process complex, since on the one side, low viscosity 
reduces shear forces, thus speeding up printing, but it can also reduce 
both resolution and accuracy \cite{webb2017parameter}. 
The dynamic viscosity (0.2-4.6 Pa $\cdot$ s) observed in alginate 
solutions represents a good compromise between 
the above criteria \cite{roopa2009characterisation}.
Further, it is worth underlining that solutions with higher alginate concentrations (up to 9\%) have been recently used for bioscaffold fabrication \cite{ILHAN20201040} with dynamic viscosity values falling in a narrow window (2.4-2.7 Pa $\cdot$ s) of the range considered in the present work. Thus, the present model is reasonable able to probe also alginate solution at higher concentration values.
On the other hand, in the actual work we focus the attention on the alginate concentrations in the range $1\%-3\%$ w/v alongside with the corresponding parameters of the Cross model reported by Roopa and Bhattacharya \cite{roopa2009characterisation}.
Sarker and Chen \cite{sarker2017modeling} have also investigated the rheology of the alginate solutions, although some parameters as $\lambda$ and $n$ were not explicitly given. 
The kinematic viscosity of the air is set $\nu_{air}=1.552\cdot 10^{-5}$ m$^2$/s, corresponding to the kinematic viscosity at 25 $^\circ$C.
The rheological parameters of the ink fluid and the air are reported in Tab.~\ref{tab:algparam}.
To model the fluid-air system we simulate two fluids with a density 
ratio $\gamma=842.0$ ($\approx$ the water/air 
density ratio at $25^{\circ}$C), while the surface tension is set at 
the typical value, $\sigma= 50.0\cdot 10^{-3}$ N/m \cite{del2005mechanisms}. 
The simulation box consist of $240\times880$ lattice nodes, the nozzle 
diameter of the channels was fixed at $d=60$ lattice nodes, and 
the same value was assigned to the distance of the nozzle by the 
deposition surface corresponding to $60$ lattice nodes. 
This configuration is characteristic of 3D bio-plotters \cite{3dbioprintgels16}. 
The system is initialised with the nozzle filled up with ink and 
located on the right side of Fig. \ref{fig:boundaries}, that also 
shows graphically the various boundary conditions used in the simulations. 
These were set periodic along the $x-$axis, while  along the $y-$ axis, 
the bottom side is a no-slip wall and  the top boundary 
outside the nozzle is set to a constant blue (air) density $\rho^B$.

\begin{table}[h]
\begin{center}
\begin{tabular}{l|c|c|c|c|c}
 & $\nu_0 \: (\text{m}^2$ s) &   $\nu_{\infty}  \: (\text{m}^2$ s)  & $\lambda$ (s) & n &  $\nu_{air} \: (m^2$ s) \\ 
\hline
  1\%-25$^{\circ}$C & $15.6\;10^{-5}$   &  $2.11\;10^{-5}$   & $3.16\;10^{-3}$ & 0.751 & $1.552\;10^{-5}$\\
  2\%-25$^{\circ}$C & $66.7\;10^{-5}$   &  $0.69\;10^{-5}$   & $5.96\;10^{-3}$ & 0.713 & $1.552\;10^{-5}$\\
  3\%-25$^{\circ}$C & $463.6\;10^{-5}$  &  $3.64\;10^{-5}$   & $62.5\;10^{-3}$ & 0.573 & $1.552\;10^{-5}$\\
  3\%-40$^{\circ}$C & $186.8\;10^{-5}$  &  $5.14\;10^{-5}$   & $10.2\;10^{-3}$ & 0.737 & $1.552\;10^{-5}$\\
 \hline
 
  & $\nu_0$  (LB) &   $\nu_{\infty}$ (LB)  & $\lambda$\:(LB) & n &  $\nu_{air}$  (LB) \\
  \hline
  1\%-25$^{\circ}$C & 0.134 & 0.0181 & $3.310\;10^{-5}$  &  0.751 & $0.01\bar{3}$ \\        
  2\%-25$^{\circ}$C & 0.430 & 0.0045 & $8.325\;10^{-5}$  &  0.713 & $0.01$  \\   
  3\%-25$^{\circ}$C & 1.991 & 0.0156 & $131.0\;10^{-5}$ &  0.573 & $0.00\bar{6}$ \\
  3\%-40$^{\circ}$C & 0.802 & 0.0221 & $21.37\;10^{-5}$ &  0.737 & $0.00\bar{6}$ 
\end{tabular} 
\end{center}
\caption{Values of rheological parameters used in simulations. For the alginate solutions the parameters were given by Roopa and Bhattacharya  \cite{roopa2009characterisation}.}
\label{tab:algparam}  
\end{table}

The ink velocity inside the nozzle in the internal reference frame is obtained from the mass flow rate given in literature \cite{sarker2017modeling}, by noting that:
\be\label{eq:fluidconv}
v_{\text{ink}}={\psi\over\rho \pi  (d/2)^2}
\ee

As reported in \cite{sarker2017modeling}, for a nozzle diameter of $0.2$ mm, 
typical value of flow rates $\psi$ are between 7.7 mg/s and 27 mg/s. 
Setting $\psi = $14 mg/s and the ink fluid velocity $v_{\text{ink}}$ by 
Eq.~\ref{eq:fluidconv}, the nozzle velocity, $v_{\text{nozzle}}$, 
was varied in range from $0.25\, v_{\text{ink}}$  to $1.75\, v_{\text{ink}}$. 
Hence, $\upsilon_{\text{inlet}}$ is computed as: 
$\vec{u}_{\text{Inlet}}=\vec{v}_{\text{ink}}+\vec{\upsilon}_{\text{nozzle}}$.

Denoting by the subscripts LB and MKS the physical observable in lattice and MKS system of units respectively, we adopted the following rules to convert the lattice units $\Delta x_{\text{LB}}, \Delta tx_{\text{LB}},\Delta m_{\text{LB}}$ in the corresponding physical quantities $\Delta x_{\text{MKS}}, \Delta tx_{\text{MKS}},\Delta m_{\text{MKS}}$.
Assuming $\Delta x_{\text{LB}}, \Delta tx_{\text{LB}},\Delta m_{\text{LB}}$ equal to one, the lattice conversion rules are:
\bea
\Delta x_{\text{MKS}}=\frac{d_{\text{MKS}}}{d_{\text{LB}}}=3.\bar{3}\cdot10^{-6} \text{ m}\label{eq:conv1}\\
\Delta m_{\text{MKS}}= \frac{(\Delta x_{\text{MKS}})^3 \rho^{air}_{\text{MKS}}}{\rho^B_{\text{LB}}}=4.39\cdot10^{-17} \text{ kg}\\
\Delta t_{\text{MKS}}= \frac{(\Delta x_{\text{MKS}})^2 \nu^B_{\text{LB}}}{\nu^B_{\text{MKS}}}= [4.77-9.55]\cdot10^{-9} \text{ s} \label{eq:visc}
\eea
In particular, if we assume the nozzle diameter $d_{\text{MKS}}=0.2 \cdot 10^{-3}$ m from Ref. \cite{sarker2017modeling} corresponding to $d_{\text{LB}}=60$ lattice nodes, the $\Delta x_{\text{MKS}}$ is determined by Eq. \ref{eq:conv1}, while $\Delta m_{\text{MKS}}$ is obtained by fixing the air density in lattice units $\rho^B_{\text{LB}}$ equal to one. 
Since the kinematic viscosity of the air is always lower than the corresponding value in the ink, $\Delta t_{\text{MKS}}$ is obtained by fixing $\tau^B$ and, thus, $\nu^B_{\text{LB}}$ is also determined by the relation $\nu^B_{LB}=c_s^2(\tau^B-0.5)$, which is inserted in Eq. \ref{eq:visc}.
In the following, $\tau^B_{\text{LB}}$ and $\nu^B_{\text{LB}}$ were fixed depending on the cases under investigations (see Tab. \ref{tab:algparam}  ), so that $\Delta t_{\text{MKS}}$ spans over the range reported in Eq. \ref{eq:visc}.   
However, $\tau^B \in [0.52,\dots,0.54]$ is taken sufficiently far from the limiting value $0.5$ in order to avoid numerical instabilities \cite{kruger2017lattice}.

\section{Results and discussion}\label{sec:04}
\begin{figure}[h!]
\includegraphics[width=0.6\hsize,clip]{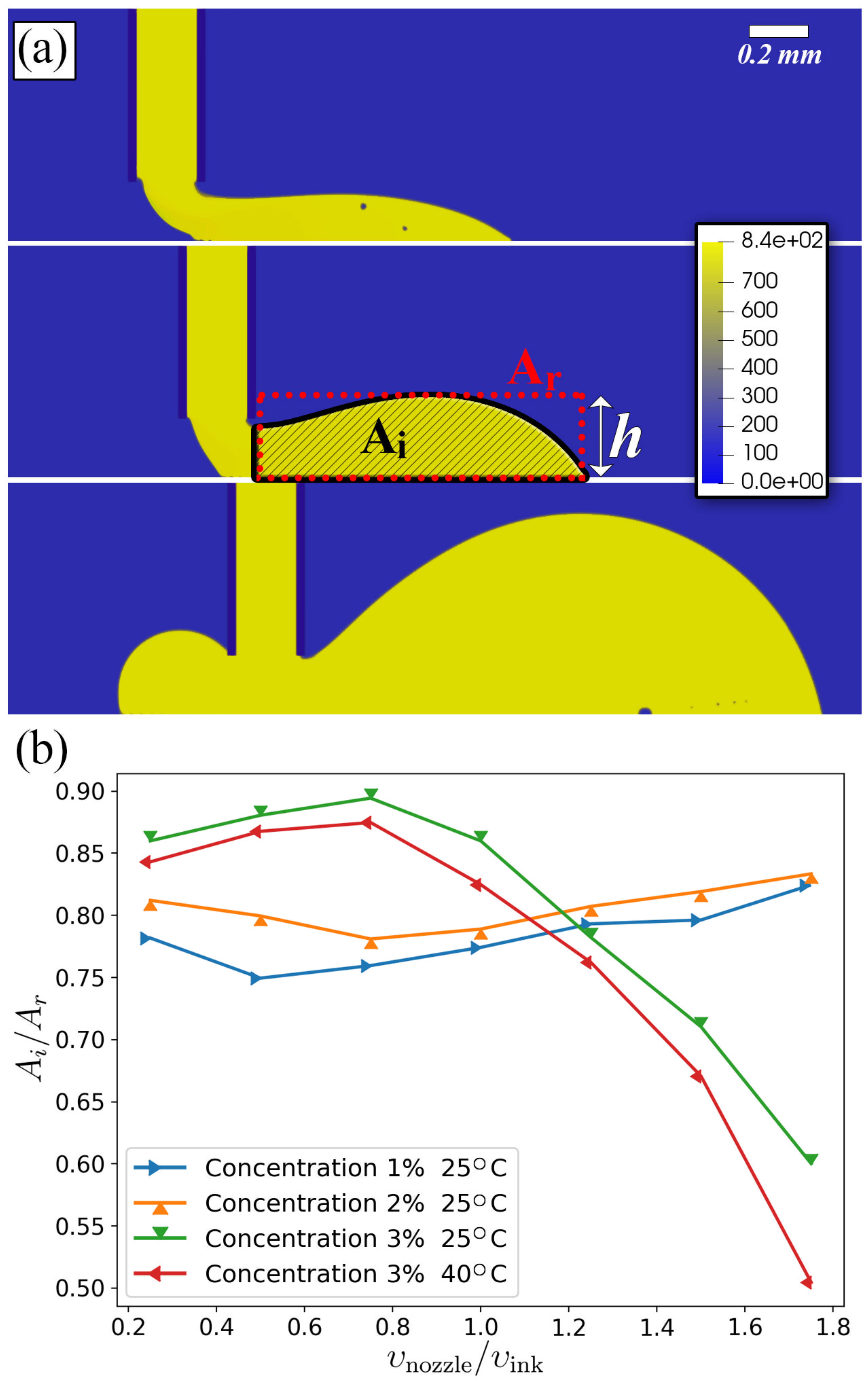}
\caption{(a) Yellow fluid densities for 1\% alginate solution. From top to bottom the nozzle velocities $\upsilon_{\text{nozzle}}$ are equal to  $1.75 \; \upsilon_{\text{ink}}$, $1.0 \; \upsilon_{\text{ink}}$, and $0.25 \; \upsilon_{\text{ink}}$. The snapshots are taken as the nozzle has travelled six times, $6d$, the nozzle diameter from the lattice position where the ink touches the deposition substrate. In the middle panel, we also sketch thread height ($h$), the area of the ink thread ($A_i$), and the area of the rectangle circumscribing the thread ($A_r$) defining the coverage ratio, $A_i/A_r$.(b) Coverage ratio obtained by different printing parameters.}
\label{Fig:per2}
\end{figure}

In 3D printing, the ultimate printability of a given prototype depends both on the printer device and on the physical properties 
of the ink fluid. In order to assess the quality of the print process with respect to tunable parameters, we introduce 
a POI following \cite{webb2017parameter}: $POI=accuracy/theoretical\,shear\,stress$.
In fact, it was found that the shear stress can be minimised by manipulating printing parameters \cite{webb2017parameter}, since 
it is proportional to the inlet pressure $p$ and inversely proportional to the nozzle diameter $d$. 
Hence, assuming the accuracy to scale inversely with the thickness (height $h$) of a single printed thread, the POI 
is written as \cite{webb2017parameter}:
\bea
POI\propto {d \over h \: p}
\eea
We include the coverage ratio (that is, the ratio of the sectional area of the ink thread, $A_i$, to the area of the rectangle circumscribing the thread, $A_r$, as in the middle panel of Fig.~\ref{Fig:per2}a) in the definition of the accuracy and obtain:
\bea
POI\propto {A_i\over A_r} \frac{d}{h \: \Delta p_{m}}
\eea
where: we consider the shear stress as proportional to the variation of the pressure in the ink, that reaches a maximum pressure variation $\Delta p_{m}$ (variation with respect to the equilibrium pressure $p_0$) outside the nozzle at completion of the deposition process; $h$ is taken as the largest height value of the thread behind the moving nozzle.

Hence, all the POI values are normalised to a reference value, in order compare the results with the different parameters set. 
We take the reference POI as the largest value, $\max_i POI_i$, corresponding to a perfect coverage ratio, $(A_i/A_r)=1$, across all the simulations. 

As a result, the normalized POI reads: $POI^{norm}_i={POI_i\over \max_i\{POI_i\}}$. 
Nonetheless, it is worth highlighting that the POI is here aimed to determine the process quality in the context of alginate-type inks used for manufacturing applications in the context, among others, of bio-scaffolds. Hence, the POI index involves the maximum pressure observed in the simulation to monitor the shear forces in the fluid. For other applications, such as manufacturing processes with polymeric inks, the shear stress can be relatively less important. Thus, other indexes of printing quality could be mainly focused on the geometrical precision rather than the shear forces in the fluid, for instance, in the contexts of nano-printing \cite{ventrici2018three} or electrode fabrication \cite{ye2018large,wei20173d}.

For all the simulations, we stopped the run as the nozzle covers six time the nozzle diameter $d$ from the lattice position where the deposited ink 
first touches the collector, thus allowing the geometry of the printed thread to be completely developed in high-resolution printing \cite{sarker2017modeling}. 

In top panel of Fig.~\ref{Fig:per2}a, a set of snapshots are reported, showing the ink mass densities map ($\rho^Y$) at the end of the simulation for three different velocity conditions of the nozzle $\upsilon_{\text{nozzle}}=\{0.25,1.0,1.75\}\upsilon_{\text{ink}}$ and alginate solution with the lowest concentration 1\%. 
The shape of the deposited ink is found to be strongly dependent on $\upsilon_{nozzle}$, overflowing beyond the travelled length of six diameters for $\upsilon_{\text{nozzle}}=0.25\;\upsilon_{\text{ink}}$ such to provide a poor printing quality. That a low dispensing speed compared to $\upsilon_{\text{ink}}$, providing a surplus of ink compared to the space spanned by the nozzle, would decrease the printing accuracy is in agreement with previous results, both experimental \cite{jin2017printability,zhang2018evaluation} and numerical \cite{agassant2019flow}. 

\begin{figure}[h!]
\includegraphics[width=0.7\hsize,clip]{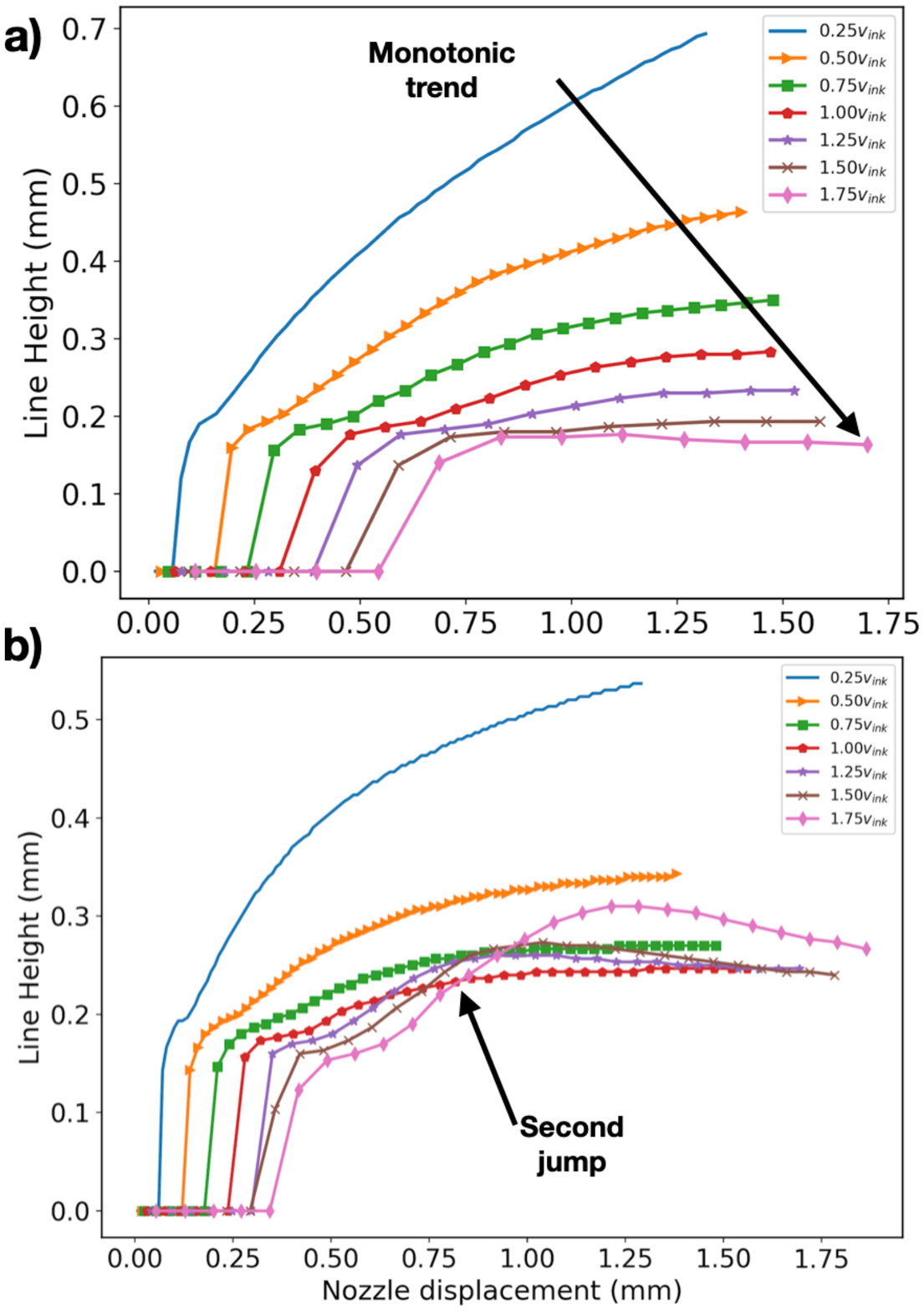}
\caption{Thread height as a function of  the nozzle displacement, for a 1\% (top panel a) and 3\% w/v (lower panel b) alginate solution. The temperature is equal to 25 $^\circ$C in both the alginate solutions.}
\label{Fig:per2height}
\end{figure}

The height of the printed thread in Fig. \ref{Fig:per2height}a is also clearly dependent on the parameter $\upsilon_{\text{nozzle}}$. In order to study the dynamic evolution, we probe the maximum height, $h$, of the fluid thread behind the moving nozzle, investigating whether a stationary condition is reached along the deposition process.

The thread height, $h$, as a function of the distance travelled by the moving nozzle is reported in Fig.~\ref{Fig:per2height}. 
In both the case of 1\% and 3\% concentrations, the height of the thread for lowest nozzle velocities, $\upsilon_{\text{nozzle}}=0.25\;\upsilon_{\text{ink}}$ 
does not show any asymptotic trend, confirming that the printing precision is deteriorated by a low dispensing velocity.  
We also observe that the asymptotic values in $h$ decrease as the nozzle velocity increases for the 1\% concentration, highlighting 
a clear, monotonic trend, which is in agreement with the experimental observations reported by Webb and Doyle \cite{webb2017parameter}. 
It is worth to highlight as the numerical results trace qualitatively the experimental trend, although the model has been implemented in the two-dimensional framework, endorsing the validity of the dimension reduction.

\begin{figure}[h!]
\includegraphics[width=0.6\hsize,clip]{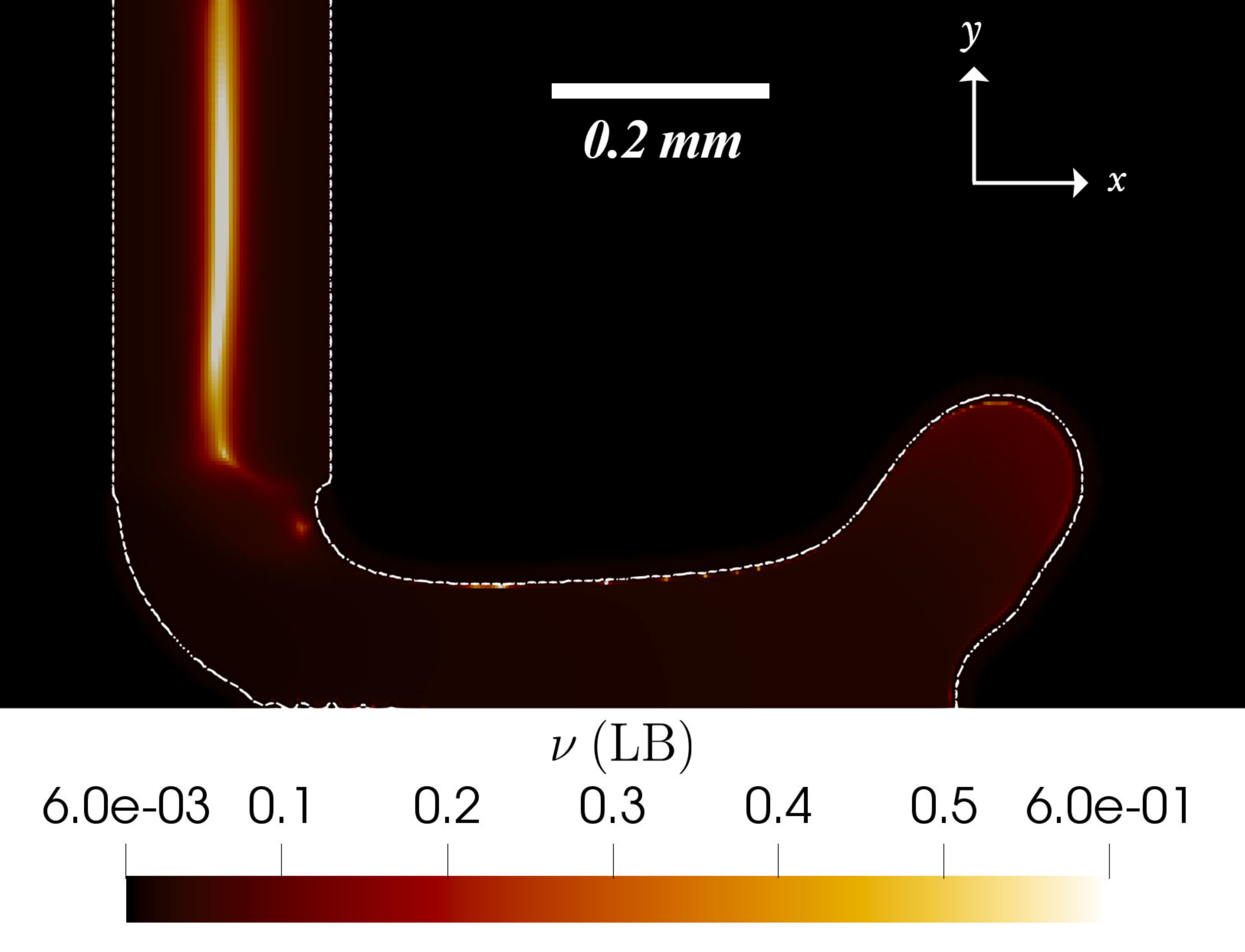}
\caption{Kinematic viscosity map of the mixture, $\nu$, for the case 3\% w/v at 25 $^\circ$C, in LB units, after the ink is deposited. The dashed white line highlights the fluid interface. The ink on the right side features higher viscosity than in the contact line with the substrate.}
\label{Fig:viscr2-3}
\end{figure}

The height evolution at high velocity and concentration 3\% w/v shows a second jump, namely the thickness obtained at higher values of $\upsilon_{\text{nozzle}}$ overcomes the value measured with lower velocities $\upsilon_{\text{nozzle}}=[0.75,1.00,1.25]\;\upsilon_{\text{ink}}$, providing 
a non monotonic trend. This suggests that, at least at high ink concentration, an optimum operating value exists for the dispensing velocity, compared 
to the ink delivery rate, which minimises the thread height. Further, the sequence of asymptotes is found to be monotonic also in the case of 2\% solutions at 25 $^\circ$C, while the sequence with concentration 3\% w/v and 40 $^\circ$C shows the same non-monotonic trend already observed at 25 $^\circ$C. The non monotonic hight trends observed for 3\% alginate concentrations (both at 25 $^{\circ}$C and 40 $^{\circ}$C) is produced by the presence of irregularities in the thread shapes as the one represented in Fig.~\ref{Fig:viscr2-3}. These irregularities manifest for large nozzle velocities and 3\% alginate concentrations are explained by viscous effect (see the discussion below) and determine also the behaviour of the coverage ratio.
The coverage ratio, reported for the four cases in Figure \ref{Fig:per2}(b), allows a similar classification, showing a maximum around $\upsilon_{\text{nozzle}}\approx  0.8 \upsilon_{\text{ink}}$ for solutions with 3\% concentration. Then, the coverage ratio decreases at higher nozzle velocity values due to the irregular shape of the deposited ink as reported in the ink density map of Fig.~\ref{Fig:per2} (panel a). The distinct irregular 
signature in the thread decreases the coverage ration at $\upsilon_{\text{nozzle}}=  1.75 \upsilon_{\text{ink}}$.

In Fig.~\ref{Fig:viscr2-3}, the fluid viscosity of the mixture is reported for the case 3\% w/v at 25 $^\circ$C after the ink is deposited.
As a first, we observe in Fig. \ref{Fig:viscr2-3} overall a low viscosity in the extruded fluid part which is the result the shear rate
enforced among the moving nozzle and the substrate.
In all the simulations, we observe a detachment point of the ink from the substrate. In particular, the shearing force produces the detachment point just after the ink reaches the substrate.
Later, the detachment point remains visible as an irregular blob in the thread (see Fig. \ref{Fig:viscr2-3}). Then, the tread reabsorbs the blob under the action of the capillary pressure.
Hence, the rheological behaviour of non-Newtonian inks play a central role in this process. 
In particular, the relation between shear-rate and the viscosity tunes the magnitude of the transmitted nozzle movements to the 
deposited ink, biasing both the thread shape and the quality of the final products.
Further, it is observed the presence of a low viscosity close to the wall of the nozzle(see Fig. \ref{Fig:viscr2-3}), which
is consistent with the Poiseuille flow as a consequence of the larger velocity gradient, $\partial{\upsilon}/\partial{x}$, close to the no-slip boundaries.
On the other hand, the viscosity profile shows a high peak in the middle of the nozzle (corresponding to the lowest velocity gradient point), 
which can be relevant for the cell viability in bio-inks. 
Indeed, in the context of cell culture applications, shear stress is essential to control cell viability, 
which may be compromised by the impact force generated by high gradient in viscosity within the nozzle channel \cite{shi2018shear,lee2018constrained}.

\begin{figure}[h!]
\includegraphics[width=0.6\hsize,clip]{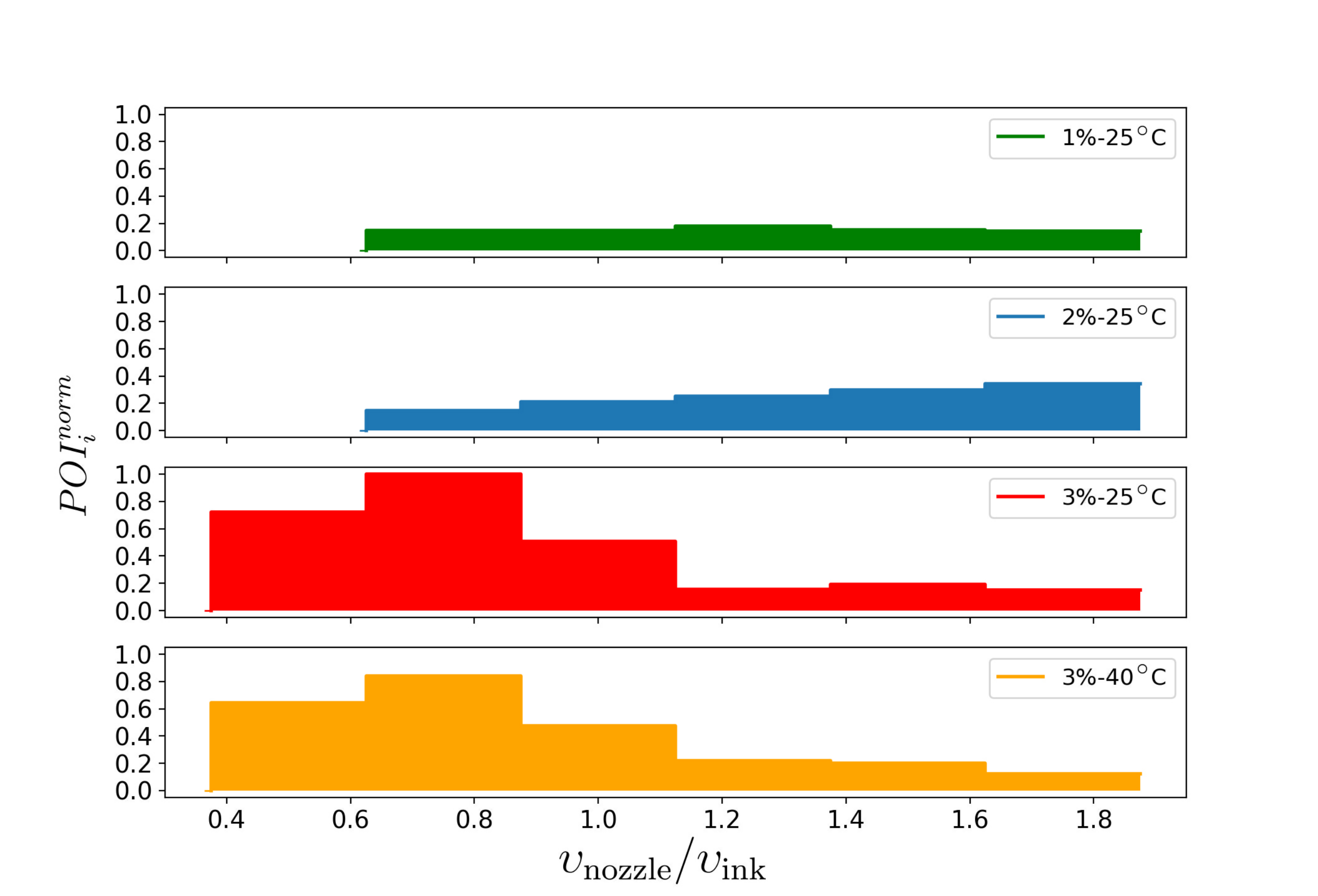}
\caption{Parameter optimization index, reported for  thread heights less than two times the nozzle diameter, for different alginate concentrations and temperatures.}
\label{Fig:poi}
\end{figure}

The $POI^{norm}_i$ values, for thread heights less than two times the nozzle diameter, are assessed and shown in Fig~\ref{Fig:poi}. Increasing the alginate concentration results in higher $POI^{norm}_i$ values, which is mainly due to the higher coverage ratios $A_i/A_r$ alongside with smaller variation in the ink pressure. 
In particular, for 3\% w/v concentrations, the ratio $\Delta p_m/p_0$ is found in the range from 0.1 to 0.2 for slow nozzle velocity $\upsilon_{\text{nozzle}}<0.8\upsilon_{\text{ink}}$. On the contrary, $\Delta p_m/\Delta p_0$ is always larger than 0.4 for 1\% and 2\% w/v and for all the nozzle velocities, providing lower $POI^{norm}_i$ values. 
Thus, the $POI^{norm}_i$ shows a peak at moderate nozzle speed in the range 0.6-0.8 $\upsilon_{\text{ink}}$ at high alginate 
concentration 3 \% w/v, due to the simultaneous concourse of high coverage ratios and small ink pressure variations.

Finally, in some cases, small porosity was noted in the ink fluid (see the upper panel of Fig.  \ref{Fig:per2}). In order to address the question, the POI values were reconsidered, taking into account the porosity.
Indeed, since $A_i$ represents the area covered by the ink, the area decreases as the porosity increases in the fluid, whenever the porous are excluded in the $A_i$ assessment.
The POI results are practically unaffected by this new definition, with $A_i$ values always changing less than 1\%. Consequently, the porosity in the trend-line does not bias the features of printed material in the present simulations, and its effect can be reasonably neglected.

\section{Conclusions}\label{sec:05}

Summarising, we have introduced a multi-component model of non-Newtonian inks through a regularised version of 
the colour gradient LB model and used it to simulate the printing process as a function of a number of design parameters. The model allows to calculate the shear stresses during the printing process of non-Newtonian inks, directly accessible by simulations, that is very important to control the cell viability in bio-inks.
The print accuracy was quantitatively analysed using the same indexes used in experimental studies \cite{webb2017parameter}. 
The impact of the pseudo-plastic rheology on the printing accuracy was investigated for a set of solutions at different alginate concentration.
Systematic investigations of processes are enabled on a broad viscosity range, providing a useful tool to probe 
the dynamics of the forces acting inside and on the ink during additive manufacturing. In real systems, shear thinning fluids are usually employed in order to favour the throughput of the device (small viscosity at high shear rates) and to obtain a stable and regular thread at the end of deposition (high viscosity at small shear rates). However the accuracy of the deposited threads is deteriorated for high viscosity alginate concentration at high print speed since these composites  favour the transmission of the inertia of the fluid impacting the substrate which produce irregularities in the threads. 

\section*{Data Availability}

Data available on request from the authors.

\begin{acknowledgments}
The research leading to these results has received funding from MIUR under the project ``3D-Phys'' (PRIN 2017PHRM8X),
and from the European Research Council under the European
Union's Horizon 2020 Framework Programme (No. FP/2014-
2020)/ERC Grant Agreement No. 739964  (\textquotedbl{}COPMAT\textquotedbl{}). We acknowledge the CINECA project DI3PRI under the ISCRA initiative, for the availability of high performance computing resources and support.
\end{acknowledgments}



%

\end{document}